\documentclass[usenatbib,useAMS,letterpaper]{mn2e}
\usepackage{graphicx}
\usepackage{ctable}
\usepackage{url}
\usepackage{times}
\usepackage{xspace}

\newcommand{\lir}{L_{\rm IR}}
\newcommand{\halpha}{H$\alpha$\xspace}

\newcommand{\msun}{~\mathrm{M_{\odot}}}
\newcommand{\msunperyr}{~\mathrm{M_{\odot} {\rm ~yr}^{-1}}}

\newcommand{\mstar}{M_{\star}}

\newcommand{\sunrise}{\textsc{sunrise}\xspace}

\newcommand{\gadgetthree}{\textsc{gadget-3}\xspace}

\newcommand{\acknowledgments}{\begin{small}\section*{Acknowledgments}\end{small}}

\title[IR luminosity as an SFR indicator]{The total infrared luminosity may significantly overestimate
the star formation rate of quenching and recently quenched galaxies}
\author[C.~C. Hayward et al.]{
\parbox[t]{\textwidth}{
Christopher C. Hayward$^{1,2}$\thanks{E-mail: christopher.hayward@h-its.org}\thanks{Moore Prize Postdoctoral Scholar in Theoretical Astrophysics},
Lauranne Lanz$^3$, Matthew L. N. Ashby$^4$, Giovanni Fazio$^4$, Lars Hernquist$^4$, Juan Rafael Mart\'{i}nez-Galarza$^4$,
Kai Noeske$^5$, Howard A. Smith$^4$, \\ Stijn Wuyts$^6$ and Andreas Zezas$^{4,7,8}$}
\vspace*{6pt} \\
$^1$TAPIR, Mailcode 350-17, California Institute of Technology, 1200 E. California Boulevard, Pasadena, CA 91125, USA \\
$^2$Heidelberger Institut f\"ur Theoretische Studien, Schloss--Wolfsbrunnenweg 35, 69118 Heidelberg, Germany \\
$^3$Infrared Processing and Analysis Center, California Institute of Technology, 1200 E. California Boulevard, Pasadena, CA 91125, USA \\
$^4$Harvard--Smithsonian Center for Astrophysics, 60 Garden Street, Cambridge, MA 02138, USA \\
$^5$Space Telescope Science Institute, 3700 San Martin Drive, Baltimore, MD 21218, USA\\
$^6$Max-Planck-Institut f\"ur extraterrestrische Physik, Postfach 1312, Giessenbachstrasse 1, 85741 Garching, Germany \\
$^7$University of Crete, Physics Department \& Institute of Theoretical \&
Computational Physics, 71003 Heraklion, Crete, Greece \\
$^8$Foundation for Research and Technology-Hellas, 71110 Heraklion, Crete, Greece}

\usepackage{hyperref}

\begin{document}

\date{Submitted to MNRAS}

\pagerange{\pageref{firstpage}--\pageref{lastpage}} \pubyear{2014}

\maketitle

\label{firstpage}

\begin{abstract}
The total infrared (IR) luminosity is very useful for estimating the star formation rate (SFR) of galaxies,
but converting the IR luminosity into an SFR relies on assumptions that do not hold for all galaxies. We
test the effectiveness of the IR luminosity as an SFR indicator by applying it to synthetic spectral energy
distributions generated from three-dimensional hydrodynamical simulations of isolated disc galaxies and
galaxy mergers. In general, the SFR inferred from the IR luminosity agrees well with the true instantaneous
SFR of the simulated galaxies. However, for the major mergers in which a strong starburst is induced,
the SFR inferred from the IR luminosity can overestimate the instantaneous SFR during the post-starburst
phase by greater than two orders of magnitude. Even though the instantaneous SFR decreases rapidly
after the starburst, the stars that were formed in the starburst can remain dust-obscured and thus produce
significant IR luminosity. Consequently, use of the IR luminosity as an SFR indicator may cause one to
conclude that post-starburst galaxies are still star-forming, whereas in reality, star formation was recently
quenched.
\end{abstract}


\section{Introduction} \label{S:intro}

The total infrared (IR) luminosity ($\lir$) of galaxies has been used to infer the star formation rate (SFR)
of galaxies for decades (see \citealt{Kennicutt:1998review} and \citealt{Kennicutt:2012} for reviews). Since
the launch of the
\textit{Herschel Space Observatory}, the IR luminosity has been often used to infer the SFRs of diverse
types of local and high-redshift galaxies \citep[e.g.][]{Rodighiero:2010,Elbaz:2011,Wuyts:2011a,
Magnelli:2012,Rosario:2012,Rosario:2013,Lee:2013}.

Using the IR luminosity as an SFR indicator has several advantages. For example, other SFR
indicators, such as recombination lines (e.g. \halpha) and ultraviolet (UV) emission, must be corrected
for dust attenuation to recover the intrinsic SFR. Correcting for dust is inherently uncertain because the
observed emission is
dominated by the less-obscured lines of sight; consequently, the dust attenuation can be underestimated.
In contrast, the IR is a promising SFR indicator precisely because of dust: in the limit of high obscuration,
essentially all of the UV-optical light from young stars is absorbed and re-emitted into the IR. Thus, the IR
luminosity should be the best SFR indicator in those situations, such as for starburst galaxies. Furthermore,
the IR luminosity can be combined
with tracers of unobscured star formation (uncorrected for dust attenuation); such combinations may
yield more accurate SFR estimates (e.g. \citealt{Kennicutt:2007,Kennicutt:2009}; \citealt*{Relano:2009};
\citealt{Wuyts:2011a,Wuyts:2011b,Reddy:2012, Lanz:2013}).

Naturally, there are also caveats to using the IR luminosity as an SFR indicator because multiple
assumptions must be made to convert the IR luminosity into an SFR (see \citealt{Kennicutt:1998review}
for a more thorough discussion). As with all SFR indicators,
the IR luminosity is sensitive to the initial mass function (IMF; e.g. \citealt{Narayanan:2012IMF}). The
sensitivity to the high end of the IMF is less severe for the
IR than for recombination lines because the IR light originates from dust that is heated by both lower-mass
stars and the massive stars that photoionise nebulae and thereby generate recombination line emission.
However, this effect can also be a disadvantage of using the IR luminosity as an SFR indicator: it is
commonly stated \citep[e.g.][]{Kennicutt:1998review} that, similar to the UV, the IR luminosity is sensitive
to the SFR averaged over the past $\sim 100$ Myr (in contrast, recombination lines are sensitive to the
SFR averaged over the past $\sim 10$ Myr). Thus, the IR
luminosity may not trace the instantaneous SFR accurately if it varies over timescales of $\la 100$ Myr.

Furthermore, dust can also be heated by older stars; if such dust heating is non-negligible, as has
been suggested by some authors (e.g. \citealt*{Sauvage:1992}; \citealt{Smith:1994};
\citealt*{Smith:1996,Walterbos:1996};
\citealt{Kennicutt:2009,Salim:2009}; \citealt{Calzetti:2010};
\citealt{Kelson:2010,Murphy:2011,Totani:2011}; \citealt{Groves:2012,
Leroy:2012}; \citealt{Fumagalli:2013,Utomo2014})
converting the IR luminosity into an SFR using a standard calibration will overestimate the true SFR.
This effect becomes increasingly important with decreasing SFR \citep{Calzetti:2010,Smith:2012,
Rowlands2014,Utomo2014}.
When the contribution of older stellar populations to dust heating is non-negligible, the implicit
averaging timescale is $>> 100$ Myr \citep{Salim:2009,Kelson:2010}.
Emission from active galactic nuclei (AGN) can also
{heat dust and thus cause the IR-inferred SFR to overestimate the true SFR.
The IR-inferred SFR can also under-predict the true SFR
if a significant fraction of the photons from young stars is not absorbed by dust or if stellar
populations younger than 100 Myr dominate the bolometric luminosity \citep[e.g.][]{Schaerer:2013,
Sklias2014}.

In this paper, we quantify the effectiveness of the IR luminosity as an SFR indicator by applying it to mock
spectral energy distributions (SEDs) calculated from three-dimensional hydrodynamical simulations of
isolated disc galaxies and galaxy mergers. Because the simulated galaxies have star formation histories
that are significantly more complex than those assumed to calibrate the $\lir$-SFR relation (i.e., a single
burst or continuous star formation history), realistically treat dust attenuation from both
stellar birth clouds and the diffuse interstellar medium (ISM), and include AGN emission, they can be used
to characterise the effects of some of the aforementioned assumptions that may cause the IR luminosity to
poorly trace the instantaneous SFR. The remainder of this work is organised as follows: in Section
\ref{S:methods}, we briefly summarise the details of the simulations. In Section \ref{S:results}, we compare
the $\lir$-inferred SFR and true instantaneous SFR and the resulting inferred and true SFR--stellar mass
($\mstar$) relations. In Section \ref{S:discussion}, we discuss some implications of our results and ways
forward.

\section{Methods} \label{S:methods}

The simulations used in this work are a subset of those used in
\citet[][hereafter L14]{Lanz2014}. We briefly describe them here, but we refer
the reader to that work for full details. We performed three-dimensional
hydrodynamical simulations of four isolated disc galaxies and
binary mergers of all possible combinations of those galaxies for a single
(non-special) orbit (thereby yielding ten merger simulations)
using the $N$-body/smoothed-particle hydrodynamics (SPH) code \gadgetthree
\citep{Springel:2005gadget}.\footnote{Recently,
various authors have highlighted issues with the standard formulation of SPH
that may cause the results of simulations performed using the
technique to be inaccurate. However, for the type of idealised simulations
used in this work, the standard form of SPH yields results that are very
similar to those of the more-accurate moving-mesh hydrodynamics technique
\citep{Hayward2014arepo}.} The progenitor disc galaxies are referred to
as M0, M1, M2, and M3, where a higher number indicates a greater mass;
the masses of the stellar discs of (M0, M1, M2, M3) are $(0.061, 0.38, 1.18, 4.22) \times 10^{10} \msun$,
and other galaxy properties, such as the gas fraction and disc scale length, are varied
such that the galaxies represent typical local galaxies \citep{Jonsson:2006,Cox:2008}.
The major (equal-mass) merger simulations used in this work are referred to as
M0M0e, M1M1e, M2M2e, and M3M3e, which correspond to mergers of two
M0, M1, M2, and M3 disc galaxies, respectively, on the \textit{e} orbit of \citet{Cox:2006}.

The simulations include gas cooling and simple models for star formation, the multiphase ISM
\citep{Springel:2003}, and black hole accretion and AGN feedback
\citep{Springel:2005feedback}. The star formation rate of the simulations is calculated
on a particle-by-particle basis: each gas particle with density greater than
a threshold density ($n_{\rm H} \sim 0.1$ cm$^{-3}$) is assigned an SFR according
to a volume-density-dependent Kennicutt--Schmidt
prescription \citep{Schmidt:1959,Kennicutt:1998}, $\rho_{\rm SFR} \propto \rho^N$;
we assume $N = 1.5$. Summing the SFRs of the individual gas particles yields
the total instantaneous SFR of the galaxy, SFR$_{\mathrm{inst}}$. In practice, star formation is
implemented by stochastically converting gas particles into star particles,
where the probability of a gas particle being converted into a star particle
depends on its SFR.

\begin{figure}
\centering
\includegraphics[width=\columnwidth]{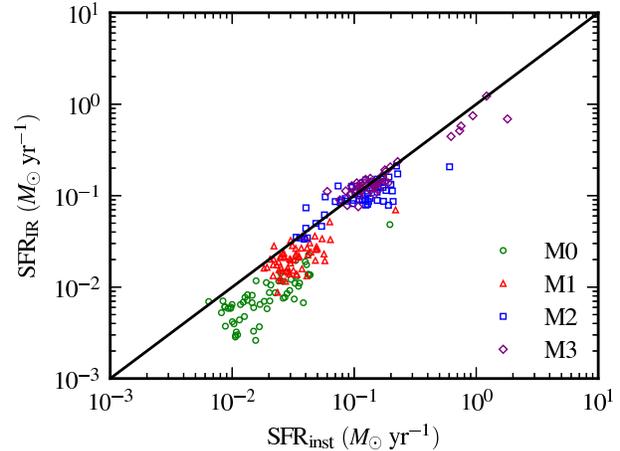}
\caption{SFR inferred from the IR luminosity vs. the instantaneous
SFR for simulated isolated disc galaxies. The points are coloured according to the
simulation from which they originate. The black line indicates equality. The inferred and actual SFR
typically agree well. For the lower-mass galaxies, a significant fraction of the photons emitted
by young stars is not absorbed; consequently, the SFR is slightly underestimated.}
\label{fig:sfr_comp_iso}
\end{figure}

To calculate mock SEDs of the simulated
galaxies, we perform three-dimensional dust radiative transfer in
post-processing at multiple times throughout each \gadgetthree simulation
using the Monte Carlo dust radiative transfer code \sunrise
\citep*{Jonsson:2006sunrise,Jonsson:2010sunrise}.
The star and black hole particles in the \gadgetthree simulations are
the sources of radiation, and the metal distribution determines the
dust distribution; we assume that 40 per cent of the metals are in
dust \citep[e.g.][]{James:2002,Sparre2014b}.
We use the default ISM model of L14, which is a conservative
choice because it exhibits less dust attenuation than the alternative model,
and we have checked that our conclusions are insensitive to whether we use the default
or alternative ISM treatments.
Given these inputs, radiative transfer is performed to calculate the
absorption, scattering, and re-emission of stellar and AGN light by dust.

These or similar simulations have been demonstrated to provide good
matches to the SEDs of local disc galaxies \citep{Jonsson:2010sunrise},
local interacting galaxies (L14), and high-redshift obscured
starbursts and AGN (\citealt{Narayanan:2009,Narayanan:2010dog,Narayanan:2010smg,Snyder:2013};
E. Roebuck et al., in preparation). Furthermore, at
different times in their evolution, they reproduce various properties of
other populations, including submillimetre galaxies
\citep[e.g.][]{Hayward:2011smg_selection,
Hayward:2012smg_bimodality,Hayward:2013number_counts},
post-starburst galaxies \citep{Snyder:2011}, and compact quiescent galaxies
\citep{Wuyts:2010}, among others. The success of the models supports our use
of them to test how well $\lir$ traces the instantaneous SFR.

\section{Results} \label{S:results}

To infer the SFR from the IR luminosity of our simulated galaxies, we first
integrate the (spatially integrated) galaxy SEDs over the wavelength range $8-1000$
\micron~to calculate $\lir$. We adopt this wavelength range following
\citet{Kennicutt:1998review}, but the precise definition (e.g. if we use only the far-IR, FIR)
is unimportant for our conclusions because the FIR luminosity dominates $\lir$.
Then, we convert $\lir$ to an SFR using the following relation:
\begin{equation} \label{eq:sfr_cal}
{\rm SFR}_{\rm IR} = 3.0 \times 10^{-37} \left(\frac{\lir}{\mathrm{W}}\right) \msunperyr,
\end{equation}
which is the \citet{Kennicutt:1998review} conversion factor converted to the \citet{Kroupa:2001}
IMF by dividing by 1.5 \citep[e.g.][]{Schiminovich:2007} . 
We have made this conversion because the \citeauthor{Kroupa:2001} IMF is assumed
when calculating the {\sc starburst99} \citep{Leitherer:1999} single-age stellar population SEDs
that are used as input for the radiative transfer calculations; thus, we must use the same IMF for
the calibration for consistency.
Also, we note that this factor is calculated by assuming continuous bursts of age 10-100 Myr, and
\citet{Kennicutt:1998review} explicitly states that strictly speaking, it is valid only for
starbursts that are less than 100 Myr old. Nevertheless, this conversion is routinely
applied to diverse galaxy types, including those for which detailed information about their SFHs
is unavailable (e.g. for high-redshift galaxies).

\begin{figure}
\centering
\includegraphics[width=\columnwidth]{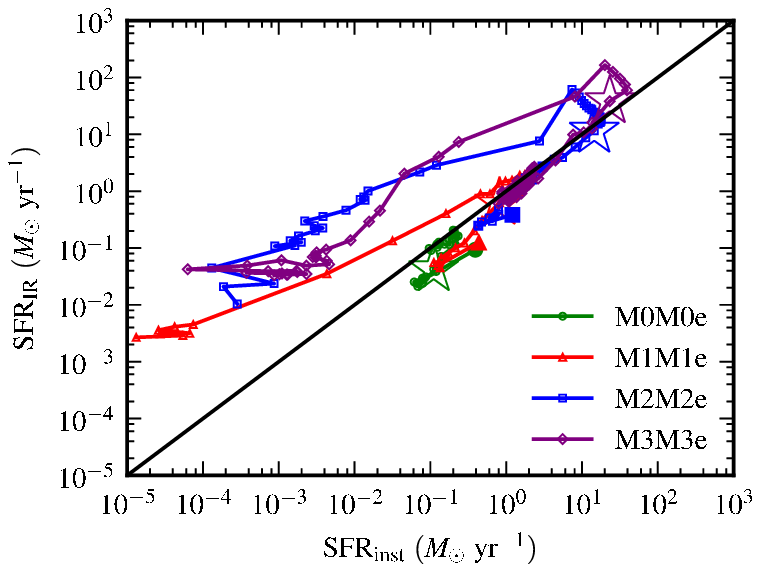} \\
\includegraphics[width=\columnwidth]{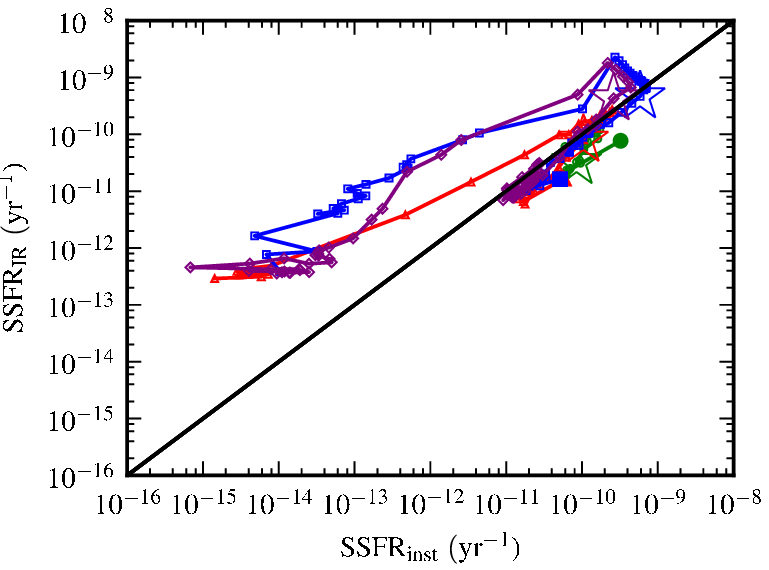}
\caption{SFR (top) and SSFR (bottom) inferred from the IR luminosity vs. the instantaneous
SFR (SSFR) for the major merger simulations.
The data points are connected to illustrate the time evolution. Near coalescence
of the black holes, which is indicated by a star, the data points are separated by 10 Myr.
At other times, the separation is 100 Myr.
The large filled symbols indicate the start of each simulation. In the simulations in which a strong starburst
is induced, the IR-inferred SFR (SSFR) can overestimate the instantaneous SFR (SSFR)
in the post-starburst phase by greater than two orders of magnitude.
Note that for fixed SSFR (of $\ga 10^{-11}$ yr$^{-1}$), the overestimation occurs only in the
post-starburst phase of the simulations. This indicates that the time evolution of the SSFR, not
just the value of the instantaneous SSFR, determines whether the IR luminosity overestimates the SSFR.}
\label{fig:sfr_comp_mergers}
\end{figure}

\subsection{Isolated disc galaxies}

Fig. \ref{fig:sfr_comp_iso} compares the SFR inferred from the IR luminosity
using Equation (\ref{eq:sfr_cal}) and the instantaneous
SFR\footnote{The `instantaneous SFR' of the simulated galaxies is
the sum of the SFRs of the individual gas particles, which are calculated based on their gas
densities and the assumed sub-resolution star formation prescription. Consequently, the instantaneous
SFR value for the simulations corresponds to an average over a timescale that is less than the maximum
time step, 5 Myr. The exact timescale is not well-defined because adaptive time steps are used.
However, the salient feature of the SFR value of the simulations is that the averaging timescale is
less than that of any of the observable SFR tracers.}
for simulated isolated disc galaxies. The data are coloured according to the
simulation from which they originate. For the major mergers, the data points are connected to
illustrate the time evolution. The start of each simulation is marked with a larger
filled symbol, and the time of coalescence of the central black holes is marked with a star.
In this and the following figures, we show the results for a single viewing angle because
the SFR is independent of viewing angle by definition, and
the variation in $\lir$ with viewing angle is $\la 10$ percent
(we note that the variation with viewing angle can be more significant in high-redshift, gas-rich mergers;
\citealt{Hayward:2012smg_bimodality}).

The inferred and true SFRs agree well for the isolated discs, although the
IR luminosity tends to underestimate the SFR of the least-massive disc galaxy
(M0). The reason for the underestimation is that this galaxy is not sufficiently
dust-obscured. Thus, a significant fraction of the light from young stars
escapes the galaxy without being absorbed by dust.

\subsection{Galaxy mergers}

Fig. \ref{fig:sfr_comp_mergers} compares the IR-inferred and true instantaneous SFR (top)
and specific SFR (SSFR $\equiv \mathrm{SFR}/\mstar$; bottom) values
for the major merger simulations. The inferred and actual (S)SFR values agree very well
for much of the mergers' evolution. However, there is an important exception: in the
simulations in which a strong starburst is induced near coalescence (all but M0M0e),
the IR-inferred (S)SFR significantly overestimates the
instantaneous (S)SFR in the post-starburst phase. This overestimation first occurs
immediately after the maximum SFR of the starburst is attained, and it can persist
for hundreds of Myr (until the end of the simulations).
At the lowest (S)SFRs, the instantaneous (S)SFR can be overestimated by greater than
two orders of magnitude.

One might expect that the SSFR determines whether older stellar populations contribute
significantly to the IR luminosity: for actively star-forming galaxies (i.e. those with high SSFR values), young
stars should dominate the dust heating, whereas for galaxies with low SSFR values, dust heating from older
stellar populations may be significant. Indeed, the bottom panel of Fig. \ref{fig:sfr_comp_mergers} indicates
that the overestimation is more significant for lower SSFR values. However, this panel also indicates that
the instantaneous value of the SSFR is not the sole determinant of how well the IR luminosity can be used
to recover the (S)SFR: for fixed SSFR (of $\ga 10^{-11}$ yr$^{-1}$, which is the minimum SSFR
value during the pre-starburst phase of the simulations), the overestimation occurs only during
the post-starburst phase of the simulations. Thus, the time evolution of the SSFR, not just the
instantaneous value, is important for determining how well the IR luminosity traces the (S)SFR.

\begin{figure}
\centering
\includegraphics[width=\columnwidth]{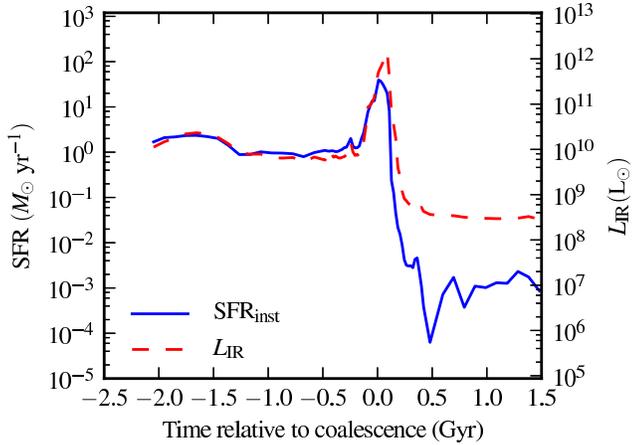}
\caption{SFR (solid blue line) and $\lir$ (dashed red line) vs. time
relative to coalescence for the most-massive major merger simulation (M3M3e).
The values on the left (SFR) and right ($\lir$) axes are related
through Eq. (\ref{eq:sfr_cal}); thus, the IR-inferred SFR can be read
off from the $\lir$ curve by referring to the left axis.
Until the peak of the starburst, $\lir$ has a similar
evolution to the SFR, and thus the IR-inferred SFR agrees with the true SFR.
However, after the starburst, $\lir$ decreases much
more gradually than the SFR because there is significant dust heating from
stars that formed during the burst; thus, $\lir$ is a poor tracer of the SFR in this regime.}
\label{fig:sfr_vs_t}
\end{figure}

To clarify the reason for the overestimation, we show the time evolution of the SFR and $\lir$ for
one of the major merger simulations, M3M3e (the most-massive simulation in our suite), in Fig.
\ref{fig:sfr_vs_t}. Near the time of final coalescence of the two galaxies, a strong starburst is induced
because tidal forces drive gas into the nuclear region
and consequently, the gas density and thus SFR increase sharply. In the starburst,
the instantaneous SFR increases by a factor of $\sim 30$ in $<200$ Myr.
Subsequently, because most of the
gas has been consumed by star formation or heated by shocks and AGN feedback
(i.e., the star formation is `quenched'), the SFR plummets by greater than four orders of magnitude.
Afterward, the SFR remains very low, and the galaxy is a passively evolving spheroid.

Until the peak of the starburst, the evolution of the SFR and $\lir$
is essentially identical. However, after the peak of the starburst, when the SFR
decreases very rapidly, this similarity no longer holds. Rather, $\lir$
declines more gradually; this difference is the origin of the aforementioned
discrepancy between the SFR inferred from $\lir$ and the instantaneous SFR.
Part of the reason for the difference is that during the starburst, the SFR varies
significantly on timescales of less than 100 Myr. Consequently, the instantaneous SFR
and SFR averaged over the past 100 Myr differ significantly at this time. Thus, one of the fundamental
assumptions of the $\lir$-SFR conversion does not hold. However, the overestimation persists and 
actually becomes more severe well after the starburst, when the instantaneous SFR is 
relatively constant on a 100-Myr timescale.

The fundamental reason for the overestimation is that although the SFR is extremely low
($\sim 10^{-3} \msunperyr$) after the starburst, a significant number
of young stars (with ages of 100s of Myr) remain. These stars are still
very luminous and obscured and hence can effectively heat the dust and thus
yield significant IR emission. Consequently, the effective averaging timescale for the IR-inferred
SFR is $>>100$ Myr during the post-starburst phase of the simulations.

In principle, the AGN can also
heat the dust and result in IR emission. By re-running the radiative
transfer calculations with the AGN disabled, we have confirmed that
dust heated by AGN is not the dominant cause of the overestimation (in all but a few snapshots of
these simulations, $\lir$ decreases by $\la 50$ per cent when the AGN emission is disabled). Nevertheless,
FIR emission from AGN-heated dust can be important in some regimes and
will be a subject of future work (L. Rosenthal et al., in preparation).

\subsection{Implications for the SFR-stellar mass relation}

\begin{figure}
\centering
\includegraphics[width=\columnwidth]{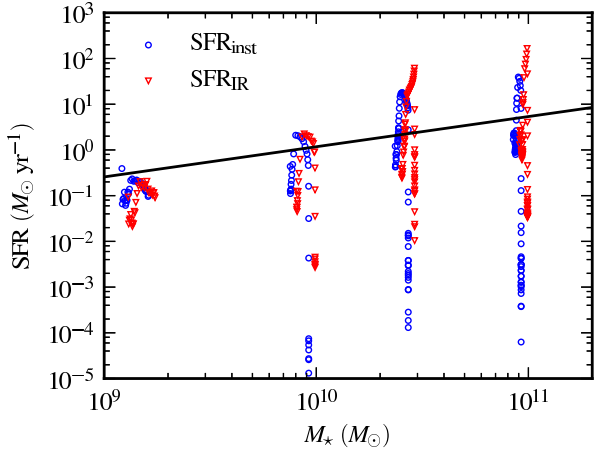}
\caption{SFR vs. $\mstar$ for the major merger simulations.
The blue circles indicate that the true instantaneous
SFR is used, whereas the red triangles indicate that the SFR value is inferred
from $\lir$. In both cases, the true $\mstar$ is used (but for clarity, the
$\mstar$ values for the red triangles have been increased by 0.03 dex).
The solid black line is the best fit to the $z \sim 0.2-0.4$ star-forming galaxies of Karim et al. (2011).
For the recently quenched galaxies, the inferred and actual locations in the
SFR--$\mstar$ plane differ significantly (i.e., there are many blue
circles that lie significantly below all of the red triangles).}
\label{fig:sfr-mstar}
\end{figure}

The overestimation demonstrated above has significant consequences for interpreting
IR observations (from e.g. \textit{Herschel} and the Atacama Large Millimeter/submillimeter Array,
ALMA) of galaxies that have
recently undergone strong starbursts. One topic for which this overestimation is potentially
significant is the SFR--$\mstar$ relation (aka `main sequence') of star-forming
galaxies \citep{Brinchmann:2004,Salim:2005,Noeske:2007a,Daddi:2007}.
Fig.~\ref{fig:sfr-mstar} shows SFR vs. $\mstar$ for the major merger simulations
for both the true instantaneous SFR and the SFR inferred from $\lir$.
Because the galaxies have low initial gas fractions (see L14 for details), the stellar mass
increases by a relatively small amount over the course of a simulation; consequently,
a given simulation primarily moves up and then down in the SFR--$\mstar$ plot as
the SFR is increased during the starburst and then quenched.

In most cases, the inferred and actual positions of
the simulated galaxies in the SFR--$\mstar$ plot agree because the IR-inferred SFR
and instantaneous SFR values are similar and the true
$\mstar$ values are used for both.\footnote{Uncertainties in
how the stellar mass is inferred from observations can also have important
consequences for the resulting SFR--$\mstar$ relation \citep[e.g.][]{Michalowski:2012}.}
However, for the post-starburst phase of the
strong-starburst mergers, this is not the case; instead, the inferred and
true locations in the SFR--$\mstar$ plot differ significantly (i.e., there are many blue
circles that lie significantly below all of the red triangles). Because the
IR-inferred SFR overestimates the true SFR in this phase, use of the IR-inferred
SFR leads to the conclusion that the recently quenched galaxies are much
closer to the SFR--$\mstar$ relation of actively star-forming galaxies (the black line in Fig. \ref{fig:sfr-mstar}) than they
actually are. This effect may also cause the apparent scatter in the SFR--$\mstar$ relation to be less
then the true scatter (see also \citealt{Schaerer:2013}),
but cosmological simulations are necessary to quantify the
effect on the scatter, which depends on the abundance of sufficiently strong
starbursts and the duration of the overestimate. Unfortunately, even state-of-the-art
cosmological simulations are not yet suitable for such an analysis
because of their relatively limited resolution \citep{Sparre2014}.

\section{Discussion} \label{S:discussion}

\subsection{Implications and scope of the overestimation}

Because of the SFR overestimation demonstrated above, when one uses $\lir$ to infer the SFR of galaxies
that have been recently quenched, inaccurate conclusions are likely.
For example, as we have demonstrated, the IR-inferred SFR may suggest
that a galaxy lies closer to the SFR--$\mstar$ relation than it actually does.
Similarly, FIR emission from AGN host galaxies
is often considered evidence that AGN do not quench star formation, which would
contradict many theoretical models \citep[e.g.][]{Springel:2005c,Hopkins:2006unified_model}.
However, the results presented above suggest that if
quenching happens rapidly (on timescales of order 100 Myr), as is suggested by
simulations, detection of FIR emission from AGN hosts does not (necessarily)
imply that star formation is ongoing in those galaxies. Other topics for
which this overestimation can have significant implications include the FIR--radio
correlation (see \citealt{Condon:1992} for a review) and the Kennicutt--Schmidt relation.

Because of the typical depths of FIR imaging, the potential bias during the post-starburst
phase that we have demonstrated is most serious for observations of low-redshift galaxies and stacking analyses
of high-redshift galaxies; i.e., for many post-starburst galaxies, SFR$_\mathrm{IR}$ may overestimate
the instantaneous SFR, but $\lir$ may be less than the detection threshold and thus
render the overestimate irrelevant.
We stress that for SFR (SSFR) values of as high as $\sim 10 \msunperyr$
($\sim 10^{-10}$ yr$^{-1}$), the instantaneous SFR can be overestimated by factors
of a few or more. This is consistent with the results of \citet{Utomo2014}, who presented observational evidence
that indicates that the SFRs inferred from a combination of UV and IR luminosities overestimate
the true values for the same (S)SFR range as in the simulations. Thus, this overestimation is
clearly relevant even for existing observational surveys and will become even more important as detection
limits are pushed lower using e.g. ALMA.

Furthermore, it is important to note that the overestimation problem is
not limited to the specific situation studied here (low-redshift galaxy mergers); rather,
whenever the SFR declines sufficiently rapidly, regardless of the absolute SFR values 
and the quenching mechanism, this overestimation should occur. Indeed, this overestimation
is also present in the post-starburst phase of simulations of $z \sim 2-3$ galaxy mergers
(see fig. 1 of \citealt{Hayward:2011smg_selection}).

\subsection{Possible solutions}

One method to correct IR-inferred SFRs for contamination from older stellar populations is to
subtract the diffuse component of the dust emission, which is likely heated predominantly by the diffuse
interstellar radiation field and thus connected with older stellar populations rather than recently formed stars
\citep[e.g.][]{Smith:1994,Smith:1996,Groves:2012,Leroy:2012,LoFaro:2013}.
However, such a correction requires spatially resolved, multi-wavelength IR data, which are typically
available only for low-redshift galaxies. It is perhaps possible to calibrate corrections using local
galaxies and apply them to high-redshift galaxies, but it is unclear that such a technique would be
effective because of the significant differences between the properties of low- and high-redshift galaxies.

Another possible approach for avoiding this overestimation is to use SFR tracers other than the IR
or a `ladder of SFR indicators' \citep{Wuyts:2011a,Wuyts:2011b}.
Radio emission is particularly promising because it is insensitive to dust obscuration and contributions
from older stellar populations \citep{Murphy:2011}. However, one must be careful to account for other
sources of radio emission.
Recombination lines are another type of SFR tracer that is insensitive to contamination from older
stellar populations. However, the most commonly used recombination line, H$\alpha$, must be corrected
for dust attenuation, and our results suggest that in the post-starburst regime, the effects of dust
are particularly nefarious: suppose that the SFR inferred from the IR luminosity is greater
than that inferred from the UV emission or recombination lines. The usual interpretation
is that dust attenuation causes the UV-optical tracers to underestimate the true SFR;
thus, the IR estimate is more accurate. However, the effect that we have demonstrated
works in the opposite direction: dust attenuation of older stars can cause the IR
to overestimate the SFR. Consequently, when UV-optical and IR SFR indicators disagree, it
is unclear a priori which is closer to the truth.

Because of this degeneracy, it is best to use broadband SED modelling, a multi-wavelength
approach that accounts for both unobscured and obscured star formation and that can place
meaningful constraints on the star formation history rather than just a single
value of the SFR that corresponds to a somewhat ill-defined time-averaged SFR.
Multiple potentially suitable tools, including {\sc magphys} \citep{daCunha:2008},
{\sc cigale} \citep{Noll:2009,Serra:2011}, and {\sc chiburst}
(\citealt{Dopita:2005,Dopita:2006c,Dopita:2006b,Groves:2008,MG2011};
R. Mart\'inez-Galarza et al., in preparation), have been developed for this task.
Application of one of these tools ({\sc magphys}) to simulated SEDs suggests that the
inferred SFRs are indeed more accurate than the IR-inferred SFRs (C. Hayward \& D. Smith,
submitted), which has also been suggested based on observational evidence \citep{Smith:2012,
Rowlands2014,Utomo2014}. The aforementioned tools and similar methods may be able to overcome the
shortcomings of simple SFR indicators such as the IR luminosity and
thus fully leverage the wealth of multiwavelength data that has become available
in recent years.

\section{Conclusions}

We have used a set of mock SEDs generated by performing dust radiative transfer
on hydrodynamical simulations of isolated disc galaxies and galaxy mergers to
investigate the effectiveness of the IR luminosity at recovering the true
instantaneous SFR of the simulated galaxies. Our principal conclusions are the
following:
\begin{enumerate}
\item For most galaxies (i.e., quiescently star-forming disc galaxies and minor mergers), the SFR inferred
from $\lir$ agrees very well with the true SFR.
\item In the simulations in which a strong starburst occurs, $\lir$
can significantly overestimate the SFR after the peak of
the starburst. The reason for this overestimation is that although the SFR
decreases rapidly from the peak, there is still significant IR emission
from dust heated by the stars formed during, and even before, the starburst.
\item The magnitude of the overestimate is greater for lower SSFR values. However, the instantaneous
SSFR value is not the sole determinant of the contribution of older stellar populations to the dust heating
because for fixed SSFR, the overestimation is more severe in the post-starburst phase of the simulations.
Thus, the time evolution of the SSFR, not just the current value, is an important determinant of how well
the IR-inferred SFR traces the true SFR.
\item This overestimation may have significant implications for e.g. the SFR-stellar mass
relation. In particular, it may cause the number of quenched galaxies and degree
to which galaxies are quenched to be underestimated.
\end{enumerate}
The results presented in this work highlight the need for caution when applying simple SFR tracers
to galaxies for which the underlying assumptions may not hold. More-sophisticated techniques, such
as SED modelling, should be used when possible, but one must also mind the caveats and potential
inherent degeneracies in these techniques.

\acknowledgments

We thank Caitlin Casey, Mattia Fumagalli, Xavier Koenig, Barry Rothberg,
Daniel Schaerer, Beverly Smith, Dan Smith, and Tomo Totani for providing useful comments
on the manuscript, and we especially thank Samir Salim and the anonymous referee for their very detailed
comments, which led to significant improvements to the manuscript.
CCH is grateful to the Klaus Tschira Foundation for financial support and
acknowledges the hospitality of the Aspen Center for Physics, which is supported
by the National Science Foundation Grant No. PHY-1066293.
HAS and LL acknowledge partial support from NASA grants NNX12AI55G and
NNX10AD68G, and JPL RSA contract 1369566. The simulations in
this paper were performed on the Odyssey cluster supported by the FAS Research
Computing Group at Harvard University. This research has made use of NASA's Astrophysics Data System.
\\

\footnotesize{
\bibliographystyle{mn2e}
\bibliography{std_citations,irsfr}
}

\label{lastpage}

\end{document}